**An extended reply to Mendez et al.: The 'extremely ancient' chromosome that still isn't**

Eran Elhaik[1*], Tatiana V. Tatarinova[2], Anatole A. Klyosov[3], and Dan Graur[4]

[1] Department of Animal and Plant Sciences, University of Sheffield, S10 2TN, UK

[2] Department of Pediatrics, Children's Hospital Los Angeles and Keck School of Medicine, University of Southern California, Los Angeles, CA 90027, USA

[3] The Academy of DNA Genealogy, Newton, MA 02459, USA

[4] Department of Biology & Biochemistry, University of Houston, Houston, TX 77204-5001, USA

Please address all correspondence to e.elhaik@sheffield.ac.uk

**Preface**

Earlier this year, we published a scathing critique of a paper by Mendez et al. (2013) in which the claim was made that a Y chromosome was 237,000-581,000 years old. Elhaik et al. (2014) also attacked a popular article in *Scientific American* by the senior author of Mendez et al. (2013), whose title was "Sex with other human species might have been the secret of *Homo sapiens*'s [sic] success" (Hammer 2013). Five of the 11 authors of Mendez et al. (2013) have now written a "rebuttal," and we were allowed to reply.

Unfortunately, our reply was censored for being "too sarcastic and inflamed." References were removed, meanings were castrated, and a dedication in the Acknowledgments was deleted. Now, that the so-called rebuttal by 45% of the authors of Mendez et al. (2013) has been published together with our vasectomized reply, we decided to make public our entire reply to the so called "rebuttal." In fact, we go one step further, and publish a version of the reply that has not even been self-censored.

A year ago, we discovered that an extremely ancient age estimate for a Y chromosomal haplotype (237,000–581,000 years ago) by Mendez et al. (2013) was based on analytical choices that consistently inflated its value. It now seems that five out of the eleven original authors of Mendez et al. (2013) disagree with our criticism of their divergence time estimates. These 45% accuse us of misunderstanding, misrepresentation, and most annoyingly "fabrication." Their rebuttal consist mainly of hypotheses that are irrefutable and, hence, unscientific, and they mostly ignore the main issues of our critique. For example, Mendez et al. (2014) claim that the time to the most common recent ancestor (TMRCA) for some human loci may exceed 1 million years. This may be true for "some human loci," even some loci on the Y chromosome, however, this claim has nothing to do with the dating in question. We have no qualms with claims of antiquity for this or that locus, as long as they are not used to promote headlines such as "Sex with other human species might have been secret of *Homo sapiens*'s [*sic*] success" (Hammer 2013). The only thing we were concerned with in Elhaik et al. (2014) was the dating of a Y chromosomal haplotype called A00.

As stated in our original criticism (Elhaik et al. 2014), estimating divergence time is not different, in principle, from estimating the time it takes two cars traveling in opposite directions at known speeds to reach a certain distance from each other. The time inferences will be overestimated if the distance between the two cars is overestimated, or if the speed of either car is underestimated. Similarly, a divergence time estimate will seem larger than the actual divergence time if the genetic distances between sequences are overestimated and/or the rates of substitution are underestimated.
Let us consider a very simple estimation model for the time of divergence,

$$t = \frac{d}{2r} \quad (1)$$

where *t* is the divergence time, *d* is the genetic distance, and *r* is the substitution rate per unit time. To overestimate *t*, one needs to overestimate *d* and/or underestimate *r*. *d* is usually estimated by dividing the number of differences between two sequences, *n*, by the length of the aligned sequences, *l*, and correcting for multiple hits and the like

$$d = \frac{n}{2l} \quad (2)$$

*d* can, thus, be overestimated by either overestimating *n* or underestimating *l*. The unit time for *r* is years. However, *r* is often derived from data on number of substitutions per generation. *r* can, thus, be overestimated by assuming that the generation time, $t_g$, is larger than it really is.

In selecting values for *d*, *r*, *n*, *l*, and $t_g$, Mendez et al. (2013) consistently and without exception chose values that led to overestimating the time of divergence. For each of these variables, Mendez et al. (2013) could have chosen from a wide range of values. However, having the conclusion of great antiquity firmly planted in their mind, they unfailingly selected values that would inflate the time of divergence estimate.

In Elhaik et al. (2014), we discussed many such choices and there is no need to refute every "refutation" in Mendez et al. (2014) as their arguments depart from the central issues we had raised. In the following we will focus on two choices left unexplained by Mendez et al. (2013).

The first choice concerns the substitution rate used in the calculation of the TMRCA. Xue et at. (2009) estimated the Y-chromosome substitution rate to be $1 \times 10^{-9}$ substitutions per nucleotide per year. Using this estimate, one can calculate divergence times of $43/240,000/10^{-9} \approx 179,000$ years and $45/180000/10^{-9} \approx 250,000$ years, for an average of 214,500 years, very similar to the TMRCA obtained using a likelihood-based method: 209,500 (95% CI: 168,000–257,400) years. Selecting an estimate based on Y chromosome for calculating the time of divergence for a Y chromosome was, however, too straightforward for Mendez et al. (2013). Instead they decided to use an autosomally derived estimate of $0.617 \times 10^{-9}$ as the mutation rate. This value is 1.6 times smaller that the estimate for Y. Unsurprisingly, Mendez et al. (2013) obtained a divergence time that is 1.6 times higher than that estimate of 290,000 to 404,000 years, with an average value of 347,000 years. More appropriate choices would have resulted in a much lower estimate. Mendez et al. (2013) other puzzling choices, such as the unprecedented 40 years for human generation time, resulted in overestimating the time of divergence by 20-130%.

The second fact that is left unexplained despite being commented on at great length by Mendez et al. (2014) concerns their highly irregular and highly questionable comparison of sequences of unequal length. In response to Mendez et al.'s (2014) allegations of "misunderstanding of population genetic theory," we challenge the authors to come up with one example in the respectable evolutionary literature in which the branches on a phylogenetic tree were estimated by using pairwise distances based on alignments of different lengths. In fact, textbooks in molecular evolution (e.g., Graur and Li 2000) specifically and strongly caution against such practices.

The most egregious accusation in Mendez et al. (2014) is one of "fabrication." In support of this claim, they took the unprecedented step of publishing exchanges of emails between Fernando Mendez and Eran Elhaik without prior approval or permission. We note that a common principle in the legal system is that telling the truth is insufficient; one should **"tell the truth, the whole truth, and nothing but the truth."** Mendez et al. (2014) published only part of the story, but not the entire truth. They omitted, for instance, the entire correspondence between Eran Elhaik (EE) and their second co-author and original discoverer of the A00 haplotype, Thomas Krahn (TK). To learn what Dr. Krahn thinks about the time estimates in Mendez et al. (2013), the missing exchange is published below with Dr. Thomas Krahn's kind permission.

TK: *"While I agree that the outrageous time estimates for A00 from Fernando [Mendez] need an urgent correction and discussion, I don't think that your argumentation yields enough weight to question the fact that Perry's Y does in fact represent an 'extremely ancient' haplogroup."*

EE: *"I am just a bit skeptical about some of their statements in the paper, that the A00 predates human and the calculation of the Y tree in their Figure 1, that doesn't sound right."*

TK: *"Yep, we were extensively discussing this issue. My numbers were more like [your] 200ky for the A00 split but Fernando [Mendez] insisted using autosomal SNP frequency data for dating, which I thought is absolute nonsense for the Y mutation mechanism. Anyhow, it was his job to do the dating maths."*

**Acknowledgements**

We thank Thomas Krahn for his comments. We dedicate this paper to two great scholars of population genetics and molecular evolution, Richard C. Lewontin and the late Allan C. Wilson, who, unfortunately, seem to have had a lesser influence on one of their students than did Hieronymus Carl Friedrich.